\documentclass[prb,aps,twocolumn,showpacs,superscriptaddress]{revtex4}

\usepackage{amsmath,bm}

\usepackage{graphicx}

\newcommand{\hk}{\bm{k}}

\newcommand{\bkl}{b_{\bm{k},\lambda}}
\newcommand{\bkld}{b_{\bm{k},\lambda}^{\dag}}

\newcommand{\bmkl}{b_{-\bm{k},\lambda}^{\dagger}}

\newcommand{\sumkl}{\sum_{\bm{k},\lambda}}

\newcommand{\wkl}{\omega_{\bm{k},\lambda}}

\DeclareMathOperator{\tr}{Tr}
\DeclareMathOperator{\re}{Re}

\begin{document}

\author{K. Roszak}
\email{katarzyna.roszak@pwr.wroc.pl}
\affiliation{Institute of Physics, Wroc{\l}aw University of
Technology, 50-370 Wroc{\l}aw, Poland}
\affiliation{Department of Condensed Matter Physics,
Faculty of Mathematics and Physics, Charles University,
12116 Prague, Czech Republic}
\author{P. Machnikowski}
\email{pawel.machnikowski@pwr.wroc.pl}
\affiliation{Institute of Physics, Wroc{\l}aw University of
Technology, 50-370 Wroc{\l}aw, Poland}

\title{Phonon-induced dephasing of singlet-triplet superpositions in
double quantum dots without spin-orbit coupling}

\begin{abstract}
We show that singlet-triplet superpositions of two-electron
spin states in a double quantum dot undergo a phonon-induced pure dephasing
which relies only on the tunnel coupling between the dots and on the
Pauli exclusion principle. As such, this dephasing process is
independent of spin-orbit coupling or hyperfine interactions. 
The physical mechanism behind the dephasing is elastic phonon 
scattering, which persists to much lower temperatures than real
phonon-induced transitions. Quantitative calculations performed for a lateral
GaAs/AlGaAs gate-defined double quantum dot yield micro-second
dephasing times at sub-Kelvin temperatures, which is consistent with
experimental observations.
\end{abstract}

\pacs{73.21.La, 03.65.Yz, 72.10.Di, 03.67.Lx}

\maketitle

\section{Introduction}

The idea of encoding quantum information in spin states of electrons confined in
quantum dots (QDs) is considered to be one of the most
promising approaches to solid-state quantum computing. The original
proposal of using single spins as quantum bits \cite{loss98} was
followed by more sophisticated concepts in which a logical qubit was
to be coded in a system of a few spins
\cite{divincenzo00,hawrylak05}. A class of proposed implementations
uses two-spin states in double quantum dots (DQDs), assigning the 
quantum logical values to singlet and triplet spin configurations
\cite{levy02,byrd02,taylor05}. State-of-the-art 
experimental techniques allow one to manipulate spin states of
two electrons in a lateral, gate-defined DQD on time scales of
hundreds of nanoseconds 
\cite{koppens06,petta05}, which has led to a renewed interest in
factors that limit spin
coherence. 

From the point of view of coherent quantum applications, two kinds of
decoherence are of interest. One is spin relaxation
(thermalization of occupations) between the selected basis states
(qubit states). The other one is dephasing of superpositions made of
these states. Such a process, often referred to as \textit{pure
  dephasing}, does not affect the occupations of the qubit basis
states, thus conserving the classical information encoded in the
system. However, by destroying quantum coherence, it degrades
the quantum bit to a classical one and hence is detrimental for
quantum information storage or processing. 
Moreover, if the system is initially prepared in a superposition of
the basis states, the pure dephasing process will lead to a decay
of this state. For a single electron
(where the qubit states are usually associated with the Zeeman
levels), this is equivalent to a decay of transverse components of the
spin \cite{merkulov02}.    

Experimental results show that spin relaxation time between the Zeeman
levels can be as long as milliseconds 
\cite{hanson03,elzerman04,kroutvar04}.
Similar values are obtained for
relaxation of singlet-triplet superpositions in two-electron systems in
double dots \cite{johnson05,hanson05}. The triplet-singlet relaxation
rate rapidly decreases already when the magnetic field reaches $0.1$~T
\cite{johnson05}. Experimentally observed spin decoherence can be
accounted for by theoretical models based on hyperfine coupling to the
magnetic moments of lattice ions 
\cite{erlingsson02,burkard99,khaetskii02,taylor07,sarkka09,coish04}.
At higher fields, when the electron
Zeeman splitting becomes much larger than the nuclear one, also phonon-assisted
processes become important for spin relaxation. The latter may either
result from an interplay of hyperfine 
and carrier-phonon interactions \cite{erlingsson02} or are mediated by
the spin-orbit coupling 
\cite{golovach04,khaetskii01,climente07a,climente07b,shen07,golovach08}.
In a nearly degenerate singlet-triplet system, the hyperfine
decoherence channel is predicted to persist even at strong magnetic
fields \cite{coish05}.

In strong contrast to these long relaxation times, the phase coherence
observed in experiments decays many orders of magnitude
faster. Time-averaged coherence times ($T_{2}^{*}$) down to tens of
nanoseconds have been observed \cite{koppens05,petta05}, while applying coherent
control techniques yields an estimate of the intrinsic value on the
order of a microsecond \cite{petta05}.

In this paper, we show theoretically that an efficient pure dephasing channel is
always present in systems of two electron spins localized in coupled
semiconductor QDs. 
This channel results solely from the charge-phonon interaction in the
presence of inter-dot tunnel coupling, and is essentially due to the
distinguishability of singlet and triplet states via Pauli-blocking of
certain transitions in the triplet case. The
key feature of this decoherence process is that it does not require
any spin-environment interaction and relies only on the
mechanisms (tunnel coupling and the Pauli principle) 
that are essential for the implementation of quantum
gates. In particular, it 
appears also in materials with negligible spin-orbit and hyperfine
couplings and is essentially independent of the magnetic field. 

The paper is organized as follows. In Sec.~\ref{sec:mechanism}, we
qualitatively explain the mechanism of dephasing. Next, in
Sec.~\ref{sec:model} we define the model to be studied and in
Sec.~\ref{sec:dynamics} we discuss its dynamics and derive the
dephasing rates. The results are presented and discussed in
Sec.~\ref{sec:results}. Section \ref{sec:concl} concludes the paper.

\section{Collisional dephasing mechanism}
\label{sec:mechanism}

Qualitatively, in the lowest energy state of the two-electron system,
each dot is occupied by a single electron. 
The spin configuration of the system may then be either
singlet or triplet. In the former case, the orbital (spatial) wave function is
symmetric and a transition to a doubly occupied state is
possible \cite{grodecka08a}. This is forbidden by Pauli exclusion in the triplet
configuration with an anti-symmetric orbital part. 
Such transitions are inefficient at sub-Kelvin
temperatures because the doubly occupied state has a higher energy and
the occupation of the required phonon states is negligible.
However, a two-phonon process is still possible, in which the absorption
of a phonon is followed by the re-emission of another one
\cite{muljarov04,muljarov05}. In such a
process, the doubly occupied state is involved only ``virtually'' and
energy conservation requires only that the two phonons have the
same energy (that is, the scattering is elastic) but this energy can
be arbitrary. 
Therefore, even at low
temperatures, phonons scatter on a DQD in the singlet state, while a
DQD in the triplet state is transparent to phonons. 
In this way, the singlet and triplet states can be
distinguished by the macroscopic environment and each scattering event
builds up correlation between the spin system and the crystal lattice. 
This distinguishability leads to
pure dephasing of any singlet--triplet superposition, in some sense
analogous to the ``collisional decoherence'' of the orbital degrees of
freedom \cite{hornberger03,machnikowski06d}. Although the intensity of
this process drops down at low temperatures because of decreasing
two-phonon spectral density at low frequencies, the temperature
dependence is only polynomial (as opposed to exponential suppression of
real transitions). As we will show, at sub-Kelvin temperatures, at
which spin coherent control experiments  
on DQDs are performed, the two-phonon process can still lead to pure dephasing
times as low as tens or hundreds of microseconds.

As the discussed process is phonon-assisted it is
characterized by a large degree of irreversibility. This follows from
the dispersive nature of the phonon reservoir which results in
correlations with a macroscopic number of uncontrollable lattice
degrees of freedom built up over the typical memory time of a phonon
bath, which is on the order of a few picoseconds.

\section{Model}
\label{sec:model}

In order to quantitatively estimate the effect of the dephasing
process discussed above we
consider two electrons in laterally coupled quantum dots interacting
with a phonon reservoir. The dots are 
considered identical and the model is restricted to the ground state
in each dot. 
This simplifying assumption strictly holds only when the Coulomb
interaction between two electrons in a single dot is smaller than the
single-particle excitation energy. Otherwise, virtual transitions to
excited states would provide another scattering channel. However, such
scattering takes place both in the singlet and triplet states and can
only yield distinguishability information due to the small
exchange-related energy shifts between the relevant states and to the
slight difference between the carrier densities in the lowest singlet and
triplet states. Thus, in general, the scattering via excited
single-particle states might lead to quantitative corrections to our
results but it should not be expected to dominate the contribution
from the channel described in Sec.~\ref{sec:mechanism}.

The Hamiltonian of the system is therefore
\begin{equation*}
H=H_{\mathrm{DQD}}+H_{\mathrm{ph}}+H_{\mathrm{int}}.
\end{equation*}
The first term describes the electrons and has the form
\begin{eqnarray}
H_{\mathrm{DQD}}
&=&-t_{1}\sum_{s}\left(  a_{\mathrm{L}s}^{\dagger}a_{\mathrm{R}s}
+\mathrm{h.c.} \right) \nonumber \\
&&+\frac{1}{2}\sum_{s,s'}\sum_{i,j,k,l}
V_{ijkl}a_{is}^{\dagger}a_{js'}^{\dagger}a_{ks'}a_{ls},
\label{H-DQD}
\end{eqnarray}
where $a_{is},a_{is}^{\dagger}$ are the electron annihilation and
creation operators with $i=\mathrm{L,R}$ denoting the left and right
dot, respectively, and $s=\,\uparrow,\downarrow$ labeling the spin
orientation. The first term in Eq.~(\ref{H-DQD}) accounts for 
single-particle inter-dot tunneling.  The second term
describes the Coulomb interaction, with $V_{ijkl}=V_{jilk}=V_{klij}=V_{lkji}$
(the wave functions may be chosen such that the matrix elements are real). 
For identical QDs the Coulomb matrix elements are also invariant
under the interchange of the dots, L$\leftrightarrow$R. 
Among the Coulomb terms, $V_{\mathrm{LRRL}}=V_{\mathrm{RLLR}}\equiv U_{1}$ and 
$V_{\mathrm{LLLL}}=V_{\mathrm{RRRR}}\equiv U_{2}$ are the
energies of the singly- and doubly-charged configurations, 
$V_{\mathrm{LRLR}}=V_{\mathrm{RLRL}}\equiv E_{\mathrm{X}}$ are
exchange energies, $V_{\mathrm{LLRR}}=V_{\mathrm{RRLL}}\equiv
t_{\mathrm{C}2}$ is the coupling between the doubly-charged
configurations, while
$V_{\mathrm{RLLL}}$ and equivalent terms account
for the coupling between the singly- and doubly-charged
configurations and will be denoted by $t_{\mathrm{C}}$.    

The Hamiltonian of the phonon reservoir is given by
\begin{equation}
\nonumber
H_{\mathrm{ph}}=\sum_{\bm{k},\lambda}
\hbar\wkl\bkld\bkl,
\end{equation}
where $\bkl,\bkld$ are phonon annihilation and creation operators for a
phonon from a branch $\lambda$ with a wave vector $\bm{k}$ and
$\hbar\wkl$ are the corresponding energies. 

Since we formulate the model in terms of states localized in
individual dots the overlap of the corresponding single particle wave
functions is negligible and the off-diagonal phonon couplings
(transferring the electrons between the dots) vanish. 
The electron-phonon interaction is therefore described by
\begin{displaymath}
H_{\mathrm{int}}=\sum_{s,i}\sumkl
F_{i}^{(\lambda)}(\hk)a_{is}^{\dagger}a_{is}(\bkl+\bmkl),
\end{displaymath}
where
\begin{displaymath}
F_{\mathrm{L/R}}^{(\lambda)}(\hk)=F^{(\lambda)}(\hk)
e^{\pm ik_x D/2}
\end{displaymath}
are coupling
constants and $D$ is the inter-dot distance. We include the deformation
potential and piezoelectric couplings. The coupling constants
for the longitudinal ($\lambda=\mathrm{l}$) and transverse
($\lambda=\mathrm{t}_{1,2}$) acoustic phonon branches are, 
respectively \cite{mahan72,roszak05b},
\begin{equation*}\label{coupling-l}
F^{(\mathrm{l})}(\bm{k}) = \sqrt{\frac{\hbar}{2\rho_{\mathrm{c}} v
\omega_{\bm{k},\mathrm{l}}}}
\left[\sigma k
- i \frac{de}{\varepsilon_{0}\varepsilon_{\mathrm{s}}}
M_{\mathrm{l}}(\hat{\bm{k}}) \right]\mathcal{F}(\bm{k}),
\end{equation*}
and
\begin{equation*}
\label{coupling-t}
F^{(\mathrm{t}_{1},\mathrm{t}_{2})}(\bm{k}) =- i
\sqrt{\frac{\hbar}{2\rho_{\mathrm{c}} v
\omega_{\bm{k},\mathrm{t}}}}
\frac{de}{\varepsilon_{0}\varepsilon_{\mathrm{s}}}
M_{\mathrm{t}_{1},\mathrm{t}_{2}}(\hat{\bm{k}}) \mathcal{F}(\bm{k}),
\end{equation*}
where $e$ denotes the
electron charge, $\rho_{\mathrm{c}}$ is the crystal density, $v$ is
the normalization volume for the phonon modes, $d$ is the
piezoelectric constant, $\varepsilon_{0}$ is the vacuum permittivity,
$\varepsilon_{\mathrm{s}}$ is the static relative dielectric constant
and $\sigma$ is the deformation potential constant. The
functions $M_{\lambda}$ depend on the orientation of the phonon wave vector
\cite{mahan72,roszak05b}. For the zinc-blende structure they are given by
\begin{equation*}\label{M}
M_{\lambda}(\hat{\bm{k}})= 2\left[
\hat{k}_{x}\hat{k}_{y}(\hat{e}_{\lambda,\bm{k}})_{z}
+\hat{k}_{y}\hat{k}_{z}(\hat{e}_{\lambda,\bm{k}})_{x}
+\hat{k}_{z}\hat{k}_{x}(\hat{e}_{\lambda,\bm{k}})_{y}
\right],
\end{equation*}
where $\hat{\bm{k}}=\bm{k}/k$ and $\hat{\bm{e}}_{\lambda,\bm{k}}$ are
unit polarization vectors. 
The form factors $\mathcal{F}(\bm{k})$ depend on the wave
function geometry and are given by
\begin{displaymath}
\mathcal{F}(\bm{k}) =
\int d^{3}\bm{r}
\psi^{*}(\bm{r})e^{i\bm{k}\cdot\bm{r}}\psi(\bm{r}),
\end{displaymath}
where $\psi(\bm{r})$ is the envelope wave function of an
electron centered at $\bm{r}=0$.

We will use the basis
composed of the three triplet states
\begin{eqnarray*}
|(1,1)T_{s}\rangle & = &
a_{\mathrm{L}s}^{\dag}a_{\mathrm{R}s}^{\dag}|0\rangle,\quad
s=\uparrow,\downarrow, \\
|(1,1)T_{0}\rangle & = & \frac{a_{\mathrm{L}\uparrow}^{\dagger}
a_{\mathrm{R}\downarrow}^{\dagger}-
a_{\mathrm{R}\uparrow}^{\dagger}
a_{\mathrm{L}\downarrow}^{\dagger}}{\sqrt{2}}|0\rangle
\end{eqnarray*}
and the three singlet states
\begin{eqnarray*}
|(\pm) S\rangle & = &\frac{|(2,0)S\rangle\pm |(0,2)S\rangle}{\sqrt{2}}\\
&=&
\frac{(a_{\mathrm{L}\uparrow}^{\dagger}a_{\mathrm{L}\downarrow}^{\dagger}\pm 
a_{\mathrm{R}\uparrow}^{\dagger}a_{\mathrm{R}\downarrow}^{\dagger})}{
\sqrt{2}}|0\rangle,\\
|(1,1)S\rangle & = & \frac{a_{\mathrm{L}\uparrow}^{\dagger}
a_{\mathrm{R}\downarrow}^{\dagger}+
a_{\mathrm{R}\uparrow}^{\dagger}
a_{\mathrm{L}\downarrow}^{\dagger}}{\sqrt{2}}|0\rangle.
\end{eqnarray*}
The triplet states and the $|(1,1)S\rangle$ singlet involve electrons occupying
separate QDs and, therefore, have lower energies than the other two
singlet states.  

The eigenstates of
the Hamiltonian $H_{\mathrm{DQD}}$ are the three triplets with the
energy $U_{1}-E_{\mathrm{X}}$, the singlet $|(-)S\rangle$ with the
energy $E_{(-)S}=U_{2}-t_{\mathrm{C}2}$, and the two states
\begin{eqnarray*}
|S_{+}\rangle&=&\frac{1}{\sqrt{1+\xi^2}}[|(+)S\rangle + \xi
|(1,1)S\rangle],\\ 
|S_{-}\rangle&=&\frac{1}{\sqrt{1+\xi^2}}[|(1,1)S\rangle -
\xi |(+)S\rangle],
\end{eqnarray*}
where 
\begin{equation}
\nonumber
\xi=\frac{2\sqrt{2}t}{U+\sqrt{U^2+8t^2}},
\end{equation}
with the
eigenenergies $E_{\pm}=\overline{E}\pm\sqrt{U^2+8t^2}/2$. Here
$\overline{E}=(U_{2}+t_{\mathrm{C}2}+U_{1}+E_{\mathrm{X}})/2$, 
$U=U_{2}+t_{\mathrm{C}2}-U_{1}-E_{\mathrm{X}}$, and
$t=\sqrt{2}(t_{\mathrm{C}}-t_1)$.  In the
weak tunneling regime, $t\ll U$, one has $\xi\ll 1$ and
$|S_+\rangle\approx |(+)S\rangle$, $|S_-\rangle\approx |(1,1)S\rangle$.
The degenerate triplet states and the singlet state $|S_-\rangle$ are
the lowest energy states.  In the following, phase decoherence of
a superposition of the $|S_{-}\rangle$ singlet state and one of the
triplet states is investigated.

Since the electron-phonon interaction conserves spin, the
singlet state $|S_{-}\rangle$ is not coupled by phonon-assisted
transitions to the triplet states. 
Calculation shows that $|S_{-}\rangle$ is also decoupled
from $|S_{+}\rangle$, so the only nonzero
off-diagonal matrix element of $H_{\mathrm{int}}$ involving $|S_{-}\rangle$ is
\begin{eqnarray*}
\lefteqn{\langle S_-|H_{\mathrm{int}}|(-)S\rangle=}\\
&&-\frac{2i\xi}{\sqrt{1+\xi^2}}
\sumkl F^{(\lambda)}(\hk)\sin\left(\frac{k_xD}{2}
\right)(\bkl+\bmkl).
\end{eqnarray*}
For this coupling, the spectral density of the phonon reservoir (as
defined, e.g., in \cite{roszak05b}) takes the
form
\begin{eqnarray}\label{R}
R(\omega)&=&\frac{1}{\hbar^2}\frac{4\xi^2}{1+\xi^2}\sumkl 
|F^{(\lambda)}(\hk)|^2 \sin^2\left(\frac{k_xD}{2}\right)\\
\nonumber
&&\times\left[(n_{\hk,\lambda}+1)\delta(\omega-\wkl)+n_{\hk}
\delta(\omega+\wkl)\right],
\end{eqnarray}
where $n_{\hk,\lambda}$ is the Bose distribution.

\section{System dynamics and dephasing rates}
\label{sec:dynamics}

The two electron system is described by the reduced density matrix
$\rho_{\mathrm{DQD}}=\tr_{\mathrm{R}}\rho$, where $\tr_{\mathrm{R}}$
denotes the partial trace over the reservoir degrees of freedom and
$\rho$ is the density matrix of the complete system. Its evolution
can be described using the time-convolutionless
(TCL) projection operator method \cite{breuer02}. For factorized
initial conditions (pure initial state), the TCL master
equation takes the form
\begin{equation}
\nonumber
\frac{d}{dt}[\mathcal{P}\rho(t)]=
\mathcal{K}(t)\mathcal{P}\rho(t),
\end{equation}
where the
projection operator $\mathcal{P}$ is defined by
\begin{displaymath}
\mathcal{P}\rho(t)=\tr_{\mathrm{R}}\rho(t)\otimes\rho_\mathrm{R},
\end{displaymath}
$\mathcal{K}(t)$ is the TCL generator, and $\rho_\mathrm{R}$ is the 
density matrix of the phonon
reservoir at the thermal equilibrium. Expanding the TCL generator up to the fourth
order yields \cite{breuer02} 
\begin{equation}
\nonumber
\mathcal{K}(t)=\sum_{n=1}^{4}\mathcal{K}_n(t),
\end{equation}
with
$\mathcal{K}_{1}(t)=\mathcal{K}_{3}(t)=0$ and
\begin{subequations}
\begin{eqnarray}\label{K24}
&&\mathcal{K}_2(t)=\int_0^t dt_1 \mathcal{PL}(t)\mathcal{L}(t_1)\mathcal{P},\\
\lefteqn{\mathcal{K}_4(t)=}\nonumber\\
&&\int_0^t dt_1 \int_0^{t_1} dt_2\int_0^{t_2} dt_3\left[\mathcal{PL}(t)\mathcal{L}(t_1)
\mathcal{L}(t_2)\mathcal{L}(t_3)\mathcal{P}\right.\nonumber\\
&&-\mathcal{PL}(t) (\mathcal{L}(t_1)
\mathcal{PL}(t_2)\mathcal{L}(t_3)+
\mathcal{L}(t_2)
\mathcal{PL}(t_1)\mathcal{L}(t_3)\nonumber\\
&&+\left.
\mathcal{L}(t_3)
\mathcal{PL}(t_1)\mathcal{L}(t_2))\mathcal{P}\right],
\end{eqnarray}
\end{subequations}
where $\mathcal{L}(t)$ is the Liouville operator,
\begin{displaymath}
\mathcal{L}(t)\rho(t')=-\frac{i}{\hbar}[H_{\mathrm{int}}(t),\rho(t')].
\end{displaymath}

The coherence between the low-energy singlet and any of the triplet
states is stored in the off-diagonal elements of the DQD density
matrix,
\begin{displaymath}
r_{i}=\langle S_{-}|\rho_{\mathrm{DQD}}|(1,1)T_{i} \rangle,
\quad i=\uparrow,\downarrow,0.
\end{displaymath}
Upon explicitly evaluating the generators in Eqs.~(\ref{K24},b), the 
evolution equation for any of these singlet-triplet off-diagonal elements
can be written in the form
\begin{displaymath}
\frac{d r_{i}}{d t}=K(t) r_{i}.
\end{displaymath}
Here, we are interested in the loss of coherence, that is, in the
evolution of the modulus of $r_{i}$. This is given by
\begin{displaymath}
\frac{d |r_{i}|}{d t}=\re K(t) |r_{i}|.
\end{displaymath}
The function $K(t)$ varies with time only for 
$t\lesssim \hbar/(k_{\mathrm{B}}T)$ (reservoir memory time), 
which is of the order of picoseconds. Since the time scales relevant
for the present discussion are many orders of magnitude longer, we can
safely use its long-time (Markovian) limit
$K\equiv K(t=+\infty)$. Then, $K$ may
be separated into two parts 
$K=K^{(1)}+K^{(2)}$. The first term is 
\begin{displaymath}
\re K^{(1)}=
-\pi(1-\epsilon)R(-\omega_{0} -\Delta E), 
\end{displaymath}
where $\hbar\omega_{0}=E_{(-)S}-E_{-}$ is the
energy difference between the two relevant singlet states,
\begin{eqnarray*}
\epsilon&=&\int^{\infty}_{-\infty}d\omega
\left[\frac{R(\omega_{0}-\omega)+R(\omega_{0}+\omega)-2R(\omega_{0})}{
2\omega^2}\right.\\
&&\left. +
\frac{R(\omega_{0}+\omega)+R(-\omega_{0}-\omega)}{
4\omega_{0}\omega}\right.\\
&&\left. -\frac{R(\omega_{0}-\omega)
+R(-\omega_{0}+\omega)}{4\omega_{0}\omega}\right],\\
\Delta E&=&\int^{\infty}_{-\infty}d\omega\left[
\frac{R(\omega_{0}+\omega)+R(-\omega_{0}-\omega)}{
4\omega_{0}\omega} \right. \\
&&\left. -\frac{R(\omega_{0}-\omega)+R(-\omega_{0}+\omega)}{4\omega_{0}\omega}
\right],
\end{eqnarray*}
and the spectral density of the phonon reservoir $R(\omega)$ is
given by Eq.~(\ref{R}). This term
describes the Fermi golden rule rate of real single-phonon transitions
between the $|S_{-}\rangle$ and 
$|(-)S\rangle$ states with fourth order corrections due to phonon-induced energy
shifts and coupling renormalizations. 
The second contribution is
\begin{eqnarray*}
\lefteqn{\re K^{(2)}=}\\
&&\frac{\pi}{2}\int_{-\infty}^{\infty}d\omega \left[
\frac{2R(\omega_{0})R(-\omega_{0})}{\omega^2}\right.\\
&&\left.
-\frac{R(\omega_{0}-\omega)R(-\omega_{0}+\omega)
+R(\omega_{0}+\omega)R(-\omega_{0}-\omega)}{\omega^2}\right.\\ 
&&\left.-\frac{R(\omega_{0}-\omega)R(-\omega_{0}+\omega)
-R(\omega_{0}+\omega)R(-\omega_{0}-\omega)}{\omega_{0}\omega}\right]
\end{eqnarray*}
and accounts for the two-phonon elastic scattering
process.  

\section{Results}
\label{sec:results}

In the calculations, DQD geometry and material parameters are taken which
correspond to lateral, gate-defined QDs made in the two-dimensional
electron gas (2DEG) of a doped GaAs/AlGaAs interface heterostructure
\cite{roszak05b,koppens06}. Two-dimensional Gaussian single electron
wave functions are used with 170~nm full width at half maximum of the
probability density.  We set $U=0.8$~meV and 
use the material parameters
$c_l=5100$~m/s, 
$c_t=2800$~m/s, $\epsilon_s=13.2$, $d=0.16$~C/m$^2$, $\sigma=-0.8$~eV,
and $\rho_c=5360$~kg/m$^3$. For simplicity, we use GaAs bulk phonon
modes. 

\begin{figure}[tb]
\begin{center}
\unitlength 1mm
\begin{picture}(85,33)(0,5)
\put(0,0){\resizebox{85mm}{!}{\includegraphics{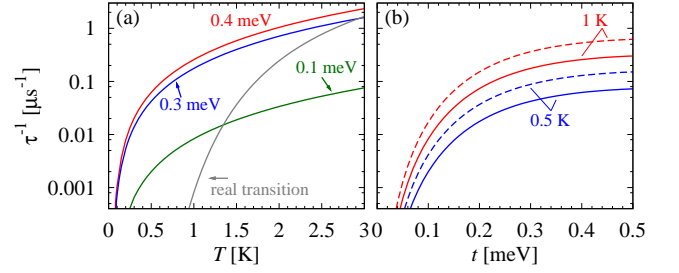}}}
\end{picture}
\end{center}
\caption{\label{fig:result} Two-phonon induced pure dephasing
  rates: (a) As a function of temperature for different tunneling 
parameters. (b) As a function of
the tunneling parameter for different temperatures.
Gray line in (a) shows the dephasing rates from single-phonon transitions
for $t=0.3$ meV. All solid lines correspond to $D=200$~nm, while the
dashed lines in (b) show results for $D=300$~nm.}
\end{figure}

The pure dephasing rates resulting from the two-phonon (scattering)
process  are shown in Fig.~\ref{fig:result}(a) as a function of
temperature. Dephasing rates due to the single-phonon assisted
transition for $t=0.3$ meV are shown in the same figure for
comparison. At low temperatures, at which experiments on DQD ensembles
are performed, the single-phonon transition is suppressed. 
At these temperatures, the dominating
decoherence mechanism is the elastic scattering. This two-phonon
process is much less influenced by 
decreasing the temperature since it involves only a virtual transition to
a higher energy (doubly charged) singlet state. The resulting pure
dephasing rates at sub-Kelvin temperatures are relatively high,
compared to experimentally achievable 
gate operation times \cite{koppens06}. 

Fig.~\ref{fig:result}(b) shows the pure dephasing rates as a function of the
coupling parameter $t$. This parameter, which is also crucial for
unitary operations on the two-qubit system, affects the dephasing rate
via the energy difference between the singlet states, $\hbar\omega_{0}$, and
via the mixing parameter $\xi$, which enters the spectral density in
Eq.~(\ref{R}). Obviously, the dephasing vanishes for uncoupled
dots. However, for non-zero coupling, the dephasing rate grows rapidly
with $t$. One can see that the dephasing rate increases as the distance
between the dots $D$ grows, which is a general feature of scattering-induced
dephasing \cite{machnikowski06d,hornberger03}.

\section{Conclusion}
\label{sec:concl}

We have shown that inter-dot tunneling and the Pauli
principle, which are necessary for two-spin quantum gate operation
in two-electron DQDs, lead to pure dephasing of singlet-triplet
superpositions by elastic phonon scattering. Although this process
does not affect occupations of the singlet and triplet states it
irreversibly destroys 
the phase coherence of quantum states by correlating the spin system
with a macroscopic number of lattice degrees of freedom which cannot
be controlled in any way.

This dephasing process does not
depend on any interactions between spins and their environment and is
therefore qualitatively different from the dephasing mechanisms
discussed so far. Unlike 
single-phonon-assisted real transitions between singlet states, which
are suppressed at low temperatures, the two-phonon elastic process
remains non-negligible in the sub-Kelvin range relevant for spin coherent 
control experiments on DQDs. For a gate-defined GaAs/InGaAs
DQD system, this process leads to dephasing rate of the order of
10~$\mu$s at $T=0.5$~K. This is
consistent with the relatively short dephasing times found in
experiments \cite{koppens06,petta05}. The elastic scattering
mechanism does not depend on magnetic field, in contrast to processes
mediated by the spin-orbit and hyperfine interactions.  

\acknowledgments

This work was supported by the Polish MNiSW under Grant
No. N~N202~1336~33 and by the Czech Science Foundation under Grant
No. 202/07/J051. 


\begin{thebibliography}{10}

\bibitem{loss98}
D. Loss and D.~P. DiVincenzo, Phys. Rev. A {\bf 57},  120  (1998).

\bibitem{divincenzo00}
D.~P. DiVincenzo, D. Bacon, J. Kempe, G. Burkard, and K.~B. Whaley, Nature {\bf
  408},  339  (2000).

\bibitem{hawrylak05}
P. Hawrylak and M. Korkusi{\'n}ski, Solid State Commun. {\bf 136},  508
  (2005).

\bibitem{levy02}
J. Levy, Phys. Rev. Lett. {\bf 89},  147902  (2002).

\bibitem{byrd02}
M.~S. Byrd and D.~A. Lidar, Phys. Rev. Lett. {\bf 89},  047901  (2002).

\bibitem{taylor05}
J.~M. Taylor, H.~A. Engel, W. D{\"u}r, A. Yacoby, C.~M. Marcus, P. Zoller, and
  M.~D. Lukin, Nature Physics {\bf 1},  177  (2005).

\bibitem{koppens06}
F.~H.~L. Koppens, C. Buizert, K.~J. Tielrooij, I.~T. Vink, K.~C. Nowack, T.
  Meunier, L.~P. Kouwenhoven, and L.~M.~K. Vandersypen, Nature {\bf 442},  766
  (2006).

\bibitem{petta05}
J.~R. Petta, A.~C. Johnson, J.~M. Taylor, E.~A. Laird, A. Yacoby, M.~D. Lukin,
  C.~M. Marcus, M.~P. Hanson, and A.~C. Gossard, Science {\bf 309},  2180
  (2005).

\bibitem{merkulov02}
I.~A. Merkulov, A.~L. Efros, and M. Rosen, Phys. Rev. B {\bf 65},  205309  (2002).

\bibitem{hanson03}
R. Hanson, B. Witkamp, L.~M.~K. Vandersypen, L.~H. {Willems van Beveren}, J.~M.
  Elzerman, and L.~P. Kouvenhoven, Phys. Rev. Lett. {\bf 91},  196802  (2003).

\bibitem{elzerman04}
J.~M. Elzerman, R. Hanson, L.~H. {Willems van Beveren}, B. Witkamp, L.~M.~K.
  Vandersypen, and L.~P. Kouwenhoven, Nature {\bf 430},  431  (2004).

\bibitem{kroutvar04}
M. Kroutvar, Y. Ducommun, D. Heiss, M. Bichler, D. Schuh, G. Abstreiter, and
  J.~J. Finley, Nature {\bf 432},  81  (2004).

\bibitem{johnson05}
A.~C. Johnson, J.~R. Petta, J.~M. Taylor, A. Yacoby, M.~D. Lukin, C.~M. Marcus,
  M.~P. Hanson, and A.~C. Gossard, Nature {\bf 435},  925  (2005).

\bibitem{hanson05}
R. Hanson, L.~H. {Willems van Beveren}, I.~T. Vink, J.~M. Elzerman, W.~J.~M.
  Naber, F.~H.~L. Koppens, L.~P. Kouwenhoven, and L.~M.~K. Vandersypen, Phys.
  Rev. Lett. {\bf 94},  196802  (2005).

\bibitem{erlingsson02}
S.~I. Erlingsson and Y.~V. Nazarov, Phys. Rev. B {\bf 66},  155327  (2002).

\bibitem{burkard99}
G. Burkard, D. Loss, and D.~P. DiVincenzo, Phys. Rev. B {\bf 59},  2070
  (1999).

\bibitem{khaetskii02}
A.~V. Khaetskii, D. Loss, and L. Glazman, Phys. Rev. Lett. {\bf 88},  186802
  (2002).

\bibitem{taylor07}
J.~M. Taylor, J.~R. Petta, A.~C. Johnson, A. Yacoby, C.~M. Marcus, and M.~D.
  Lukin, Phys. Rev. B {\bf 76},  035315  (2007).

\bibitem{sarkka09}
J. S\"{a}rkk\"{a} and A. Harju, Phys. Rev. B {\bf 79},  085313  (2009).

\bibitem{coish04}
W.~A. Coish and D. Loss, Phys. Rev. B {\bf 70},  195340  (2004).

\bibitem{golovach04}
V.~N. Golovach, A. Khaetskii, and D. Loss, Phys. Rev. Lett. {\bf 93},  016601
  (2004).

\bibitem{khaetskii01}
A.~V. Khaetskii and Y.~V. Nazarov, Phys. Rev. B {\bf 64},  125316  (2001).

\bibitem{climente07a}
J.~I. Climente, A. Bertoni, G. Goldoni, M. Rontani, and E. Molinari, Phys. Rev.
  B {\bf 75},  081303(R)  (2007).

\bibitem{climente07b}
J.~I. Climente, A. Bertoni, G. Goldoni, M. Rontani, and E. Molinari, Phys. Rev.
  B {\bf 76},  085305  (2007).

\bibitem{shen07}
K. Shen and M.~W. Wu, Phys. Rev. B {\bf 76},  235313  (2007).

\bibitem{golovach08}
V.~N. Golovach, A. Khaetskii, and D. Loss, Phys. Rev. B {\bf 77},  045328
  (2008).

\bibitem{coish05}
W.~A. Coish and D. Loss, Phys. Rev. B {\bf 72},  125337  (2005).

\bibitem{koppens05}
F.~H.~L. Koppens, J.~A. Folk, J.~M. Elzerman, R. Hanson, L.~H. {Willems van
  Beveren}, I.~T. Vink, H.~P. Tranitz, W. Wegscheider, L.~P. Kouwenhoven, and
  L.~M.~K. Vandersypen, Science {\bf 309},  1346  (2005).

\bibitem{grodecka08a}
A. Grodecka, P. Machnikowski, and J. F{\"o}rstner, Phys. Rev. B {\bf 78},
  085302  (2008).

\bibitem{muljarov04}
E.~A. Muljarov and R. Zimmermann, Phys. Rev. Lett. {\bf 93},  237401  (2004).

\bibitem{muljarov05}
E.~A. Muljarov, T. Takagahara, and R. Zimmermann, Phys. Rev. Lett. {\bf 95},
  177405  (2005).

\bibitem{hornberger03}
K. Hornberger and J.~E. Sipe, Phys. Rev. A {\bf 68},  012105  (2003).

\bibitem{machnikowski06d}
P. Machnikowski, Phys. Rev. Lett. {\bf 96},  140405  (2006).

\bibitem{mahan72}
G.~D. Mahan,  in {\em Polarons in Ionic Crystals and Polar Semiconductors},
  edited by J.~T. Devreese (North-Holland, Amsterdam, 1972).

\bibitem{roszak05b}
K. Roszak, A. Grodecka, P. Machnikowski, and T. Kuhn, Phys. Rev. B {\bf 71},
  195333  (2005).

\bibitem{breuer02}
H.-P. Breuer and F. Petruccione, {\em The Theory of Open Quantum Systems}
  (Oxford University Press, Oxford, 2002).

\end{thebibliography}

\end{document}